\documentclass[twocolumn,pra,aps,superscriptaddress]{revtex4-2}

\usepackage[latin9]{inputenc}
\usepackage{mathtools}
\usepackage{amsbsy}
\usepackage{amssymb}
\usepackage{amstext}
\usepackage{bm}
\usepackage{xcolor}
\definecolor{darkblue}{rgb}{0, 0, 1}
\newcommand{\RN}[1]{%
	\textup{\uppercase\expandafter{\romannumeral#1}}%
}
\usepackage[unicode=true,pdfusetitle,
bookmarks=false,
breaklinks=false,pdfborder={0 0 1},backref=false,colorlinks=true,allcolors=darkblue]
{hyperref}
\hypersetup{
	bookmarksnumbered=false,bookmarksopen=false}
\makeatletter

\@ifundefined{textcolor}{}
{%
	\definecolor{BLACK}{gray}{0}
	\definecolor{WHITE}{gray}{1}
	\definecolor{RED}{rgb}{1,0,0}
	\definecolor{GREEN}{rgb}{0,1,0}
	\definecolor{BLUE}{rgb}{0,0,1}
	\definecolor{CYAN}{cmyk}{1,0,0,0}
	\definecolor{MAGENTA}{cmyk}{0,1,0,0}
	\definecolor{YELLOW}{cmyk}{0,0,1,0}
}

\newcommand{\beq}{\begin{equation}}
\newcommand{\eeq}{\end{equation}}
\newcommand{\beqa}{\begin{eqnarray}}
\newcommand{\eeqa}{\end{eqnarray}}

\usepackage{amsmath}
\usepackage{graphicx}
\usepackage{amssymb}
\usepackage{txfonts,color}
\makeatother

\begin{document}
\title{Dropout is all you need: robust two-qubit gate with reinforcement learning}

\author{Tian-Niu Xu}
\affiliation{College of Information Science and Engineering, Qilu Normal University, Jinan 250013, China}

\author{Yongcheng Ding}
\affiliation{Department of Physical Chemistry, University of the Basque Country UPV/EHU, Apartado 644, 48080 Bilbao, Spain}
\affiliation{Department of Physics, Shanghai University, 200444 Shanghai, China}

\author{Jos\'e D. Mart\'in-Guerrero}
\email{jose.d.martin@uv.es}
\affiliation{IDAL, Electronic Engineering Department,  ETSE-UV, University of Valencia, Avgda. Universitat s/n, 46100 Burjassot, Valencia, Spain}
\affiliation{Valencian Graduate School and Research Network of Artificial Intelligence (ValgrAI), Valencia, Spain}

\author{Xi Chen}
\email{xi.chen@ehu.eus}
\affiliation{Department of Physical Chemistry, University of the Basque Country UPV/EHU, Apartado 644, 48080 Bilbao, Spain}	
\affiliation{EHU Quantum Center, University of the Basque Country UPV/EHU, 48940 Leioa, Spain}

\date{\today}

\begin{abstract}	
	
In the realm of quantum control, reinforcement learning, a prominent branch of machine learning, emerges as a competitive candidate for computer-assisted optimal design for experiments. This study investigates the extent to which guidance from human experts is necessary for the effective implementation of reinforcement learning in designing quantum control protocols. Specifically, we focus on the engineering of a robust two-qubit gate within a nuclear magnetic resonance system, utilizing a combination of analytical solutions as prior knowledge and techniques from the field of computer science. Through extensive benchmarking of different models, we identify dropout, a widely-used method for mitigating overfitting in machine learning, as an especially robust approach. Our findings demonstrate the potential of incorporating computer science concepts to propel the development of advanced quantum technologies.
\end{abstract}

\maketitle

\section{Introduction}

With the advancements in algorithms, computer-designed experiments~\cite{krenn2016,krenn2020} have disrupted the traditional notion that human experts are solely responsible for proposing new experiments. This raises the fundamental question: How much information must human experts provide to algorithms for efficient setup searching? This question becomes particularly pertinent when considering computer-assisted experiment design as a simplified version, where one is aware of the experimental setup but seeks specific protocols to achieve desired goals. In this scenario, reinforcement learning (RL) emerges as a natural choice, allowing agents to explore solutions by interacting with the environment. RL~\cite{francois2018} has seen significant applications in studying physics problems over the past decade~\cite{martin2021}, primarily focusing on quantum control problems~\cite{yao2021, bukov2018, porotti2019, niu2019, dalgaard2020, zhang2018, wu2019, ostaszewski2019, jiang2022}. Meanwhile, its collaboration with artificial neural network (ANN) has been employed to solve pulse design for quantum state preparation~\cite{zhang2019,haug2020,messikh2022}, gate operation~\cite{an2019,ding2021,ai2022,dong2023}, and the quantum Szilard engine~\cite{sordal2019}. Furthermore, RL has been utilized in information retrieval, controlling the measurement process in quantum metrology, and extracting real-time quantum information for closed-loop quantum control~\cite{borah2021,ding2023}. Additionally, quantum RL has also been developed by replacing the classical models with quantum system due to the model-free nature~\cite{dong2008}. 

To explore the potential of RL in the presence of limited physical knowledge, a targeted selection of a quantum control task is imperative. In light of compelling motivations, our focus is directed towards the design of robust quantum gates. Simulating the dynamics of quantum systems for building the RL environments becomes exponentially complex as the system size increases. However, by numerically calculating the propagator of a Hamiltonian with only a few energy levels per episode, we can present fair comparisons among various models with afforable computational resources. From the physical perspective, numerous approaches to achieving robustness have been proposed, such as adiabatic quantum control~\cite{kral2007}, pulse-shaping engineering~\cite{steffen2007, barnes2012, daems2013}, composite pulses~\cite{brown2004, torosov2011,rong2015,wu2023,shi2023}, and shortcuts to adiabaticity (STA)~\cite{guery2019,chen2010}. Particularly, geometric quantum computing, which leverages topological property to cancel the effects of systematic errors, has shown promise across various quantum systems~\cite{leibfried2003,zhu2003,ota2009,abdumalikov2013, zu2014, zhang2020}, and has been combined with machine learning (ML) methods~\cite{liu2019,dridi2020,mao2023}. It is indeed a well-studied example that provides sufficient information for encoding in the environment to explore the necessity of physics knowledge. Based on our research background, we believe it is possible to design robust quantum gates without considering explicit methods, e.g., accumulating geometric phase or canceling error sensitivity.

This work is organized as follows. In Section \ref{model}, we present an analytical solution for the geometric two-qubit gate, serving as an example of robust quantum gate design based on physical knowledge. In Section \ref{DRL}, we propose a RL model that encodes the pertinent information, including operation time and tunable parameter ranges. Section \ref{Numerical}  and \ref{discussion} focuse on benchmarking the performances of RL models trained under different settings and methods from computer science, such as perturbation on nodes or dropout.  We find out that the dropout itself leads the model to robustness against systematic errors without guidance from human experts. Finally, we provide concluding remarks in Section \ref{conclusion}, discussing potential avenues for further research and the implications of our findings for practical quantum information and quantum computing.

\section{Theoretical model}

\label{model}

\begin{figure}
\includegraphics[width=8.6cm]{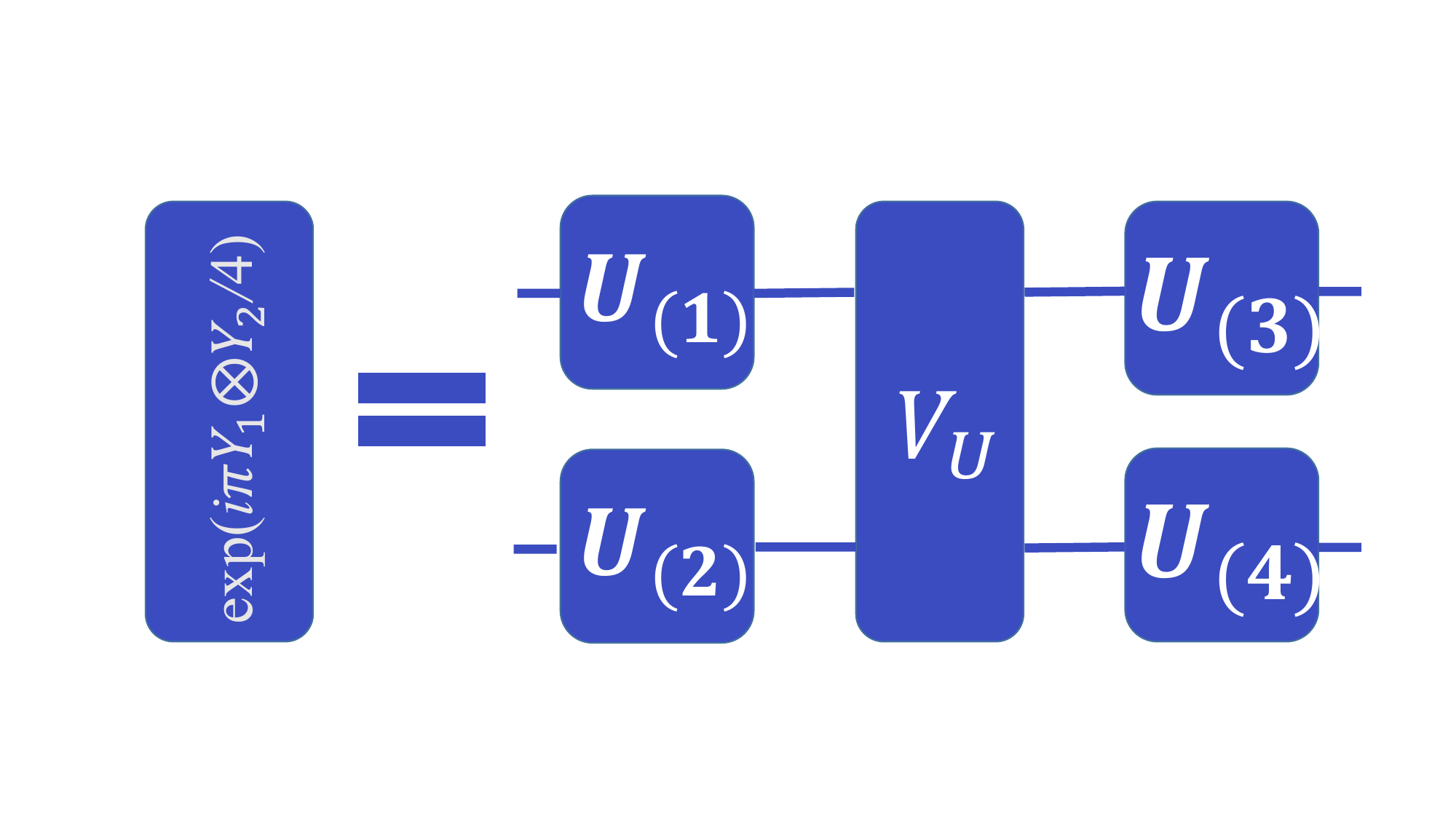}
\caption{\label{fig:scheme} Decomposition of a robust two-qubit entangling gate, $R_{YY}(\pi/4)$, highlighting the topological protection mechanism against systematic errors. The entangling gate is decomposed into universal geometric qubit gates, which are analytically solvable by canceling dynamical phases. The gate operation requires a total time of $\max(T(U_{(1)}),T(U_{(2)})) + T(V_U) + \max(T(U_{(3)}),T(U_{(4)}))$. Information gained from the analytical solution, such as the operation time, switching time, and form of Hamiltonian, is utilized to design the environment for investigating the capabilities of deep reinforcement learning.}
\end{figure}

Before going for RL, we have to specify our task and present an analytical solution if possible. Here we look for the robust entangling gate $R_{YY}(\pi/4)$. A well-known approach is the geometric gate, whose robustness mechanism is induced by topological protection. As a perfect entangling gate that converts separable states into maximally entangled state, it can serve the same critical character as CNOT does in the construction of universal gate set for quantum computing. The gate exploits the geometric phase acquired by a quantum state during its cyclic evolution in parameter space, thereby offering inherent resilience against systematic errors on the controlling parameters. The design of Hamiltonians plays a crucial role in shaping the geometric properties of the system. Correspondingly, we write down the general Hamiltonian ($\hbar=1$) as
\begin{eqnarray}
\label{eqn:H}
H &=& H_1\otimes I_2 + I_1\otimes H_2 + \frac{J}{2}Z_1\otimes Z_2,\nonumber\\
H_{1,2} &=& \frac{1}{2}[\Omega\cos(\omega t) X_{1,2}+\Omega\sin(\omega t) Y_{1,2} + \Delta Z_{1,2}],
\end{eqnarray}
where $\Omega_i$, $\Delta_i$, $X_i,~Y_i,~Z_i$ are the Rabi frequency, detuning, and Pauli operators on the $i$-th qubit, respectively, and $J$ is the exchange energy between two qubits. As a nonadiabatic approach, we employ the Lewis-Riesenfeld theory and inverse engineering for fast and robust gate operation. For obtaining the dynamical invariant, we intentionally turn off the local pulses on the first qubit, and simplify the two-qubit Hamiltonian to two parts
\begin{equation}
\label{eq:Hpm}
H^\pm(t) = \frac{1}{2}[\Omega\cos(\omega t)G_x^\pm + \Omega\sin(\omega t)G_y^\pm+\Delta^\pm G_z^\pm],
\end{equation}
where $G_i^\pm=(I_1\pm Z_1)/2\otimes(i\in\{X,Y,Z\})$ are effective Pauli operators and $\Delta^\pm=\Delta\pm J$ is the effective detuning. By mimicking the invariant of a two-level Hamiltonian, we have the dynamical invariant of $H^\pm$ as
\begin{equation}
I^\pm(t) = \Omega\cos(\omega t) G_x^\pm + \Omega\sin(\omega t) G_y^\pm + (\Delta^\pm-\omega)G_z^\pm.
\end{equation}
The instantaneous eigenstates of the Hamiltonian $H^\pm$ is in the superposition of the eigenvectors of the invariant $I^\pm$ as $|\Psi(t)\rangle=\sum_nc_ne^{i\alpha_n(t)}|\phi(t)\rangle$, where $\alpha_n=\int_0^Tdt'\langle\phi_n(t')|i\partial_{t'}-H(t')|\phi_n(t')\rangle$ is the Lewis-Riesenfeld phases that carry both geometric phases and dynamical phases. By canceling the dynamical phases $\gamma_n^d=-\int_0^Tdt'\langle\phi_n(t')|H(t')|\phi_n(t')\rangle$, the geometric gate emerges as
\begin{equation}
U(t,0)=\sum_n e^{i\alpha_n(t)}|\phi_n(t)\rangle\langle\phi_n(0)|,
\end{equation}
which is governed by geometric phases only after operation time of $T$. To study the Lewis Riesenfled phases, we analyze the eigensystem of $I^\pm$:
\begin{equation}
	E_\pm^+ = \pm \lambda^+ ,\quad
	|\phi_\pm^+(t)\rangle=
	\left(
	\begin{array}{c}
		\cos\theta_\pm^+ e^{-i\omega t} \\
		-\sin\theta_\pm^+ \\
		0\\
		0\\
	\end{array}
	\right),
\end{equation}
\begin{equation}
	E_\pm^- = \pm \lambda^- ,\quad
	|\phi_\pm^-(t)\rangle=
	\left(
	\begin{array}{c}
		0\\
		0\\
		\cos\theta_\pm^- e^{-i\omega t} \\
		-\sin\theta_\pm^- \\
	\end{array}
	\right),
\end{equation}
where $\lambda^\pm = \sqrt{(\Delta^\pm-\omega)^2+\Omega^2}$ are the eigenvalues, $\alpha^\pm_\pm = (\lambda^\pm \mp \omega)t/2$ are the Lewis-Riesenfeld phases, $\sin\theta_\pm^\pm = 1/\sqrt{{\xi_\pm^\pm}^2+1}$, 
$\cos\theta_\pm^\pm = \xi_\pm^\pm/\sqrt{{\xi_\pm^\pm}^2+1}$, and $\xi_\pm^\pm = \Omega/(\Delta^\pm \mp\lambda^\pm-\omega)$. Accordingly, we derive the setting $\Delta=\omega/2,~\Omega=\pm\sqrt{\omega^2-4J^2}/2$ in the nonadiabatic regime, and the two-qubit geometric gate
\begin{widetext}
	\begin{equation}\label{eq:VU}
		\begin{split}
			V_U &= e^{i\gamma^+_\pm}|\phi^+_\pm(0)\rangle\langle\phi^+_\pm(0)| + e^{i\gamma^-_\pm}|\phi^-_\pm(0)\rangle\langle\phi^-_\pm(0)|,\\
			&= \left(
			\begin{array}{cccc}
				-\cos(\pi a_-)-ia_-\sin(\pi a_-) & ia_+\sin(\pi a_-) & 0 & 0 \\
				ia_+\sin(\pi a_-)  & -\cos(\pi a_-)+ia_-\sin(\pi a_-) & 0 & 0 \\
				0 & 0 & -\cos(\pi a_+)-ia_+\sin(\pi a_+) & ia_-\sin(\pi a_+)\\
				0 & 0 & ia_-\sin(\pi a_+) & -\cos(\pi a_+)+ia_+\sin(\pi a_+)
			\end{array}\right),\\
			&= -e^{i\pi a_-\left(-a_+G_x^++a_-G_z^+\right)}e^{i\pi a_+\left(-a_-G_x^-+a_+G_z^-\right)},
		\end{split}
	\end{equation}
\end{widetext}
where $a_\pm=\sqrt{J/\omega\pm1/2}$. To ensure that the gate $V_U$ allows maximum entanglement, we calculate the singular values $D_\pm$ of the matrix $D$, which its elements are $D_{ij}=\text{Tr}(V_Ui\otimes j)/4$ with $i,j\in\{I,X,Y,Z\}$, and ensure that it equals to $D_\pm^{\text{CNOT}}=\sqrt{1/2}$ or other perfect entanglers. Accordingly, we find the parameter setting $J/\omega=\pm0.3187$ that corresponds to the gate time of $V_U$ as $T=2\pi/\omega$. With Cartan decomposition, one can verify that the entangling part of $V_U$ is exactly $\exp(i\pi Y_1Y_2/4)$. For the standard $R_{YY}(\pi/4)$ gate, we have local single qubit geometric gates on each qubit before and after $V_U$ as shown in Fig.~\ref{fig:scheme}, which are all robust against systematic errors on local Rabi frequency and detuning. The method for designing single qubit gate is similar to the two-qubit gate. The dynamical invariant of $H_{1,2}$ shares the same structure as $I^\pm$ but with effective operators $G_i^\pm$ and effective detuning $\Delta^\pm$ replaced by the standard version $X_i,~Y_i,~Z_i$ and $\Delta$, respectively. We also have the nonadiabatic condition for canceling the dynamical phases as $\Omega^2+\Delta(\Delta-\omega)=0$, yielding the parameterized single qubit gate of gate time
\begin{equation}
\label{eq:Ui}
U_i(\beta_j)=-\exp[i\pi\sin\beta_j(-\cos\beta_j X_i+\sin\beta_j Z_i)],
\end{equation}
where $\cos^2\beta_j=\Delta/\omega_j$ with the gate time $T=2\pi/\omega_j$. By a sequence of $\beta_j$ on the $i$-th qubit gate, one achieves the universal single qubit rotation gate with three Euler angles. For simplicity, we minimize the Frobenius norm $||U-R_{YY}(\pi/4)||$ between the matrix expression of the target gate $R_{YY}(\pi/4)=\exp(-i\pi Y_1\otimes Y_2/4)$ and $U=(U_{(3)}\otimes U_{(4)})V_U(U_{(1)}\otimes U_{(2)})$, resulting in
\begin{eqnarray}
U_{(1)} &=& U_1(0.13)U_1(0.91)U_1(0.29)U_1(0.52),\\
U_{(2)} &=& U_2(0.46)U_2(0.31)U_2(0.90)U_2(0.3)U_2(0.69)\nonumber \\&~& U_2(0.23)U_2(0.48),\\
U_{(3)} &=& U_1(0.24)U_1(0.56)U_1(0.29)U_1(0.24)U_1(0.81)\nonumber \\&~&U_1(0.29)U_1(0.81),\\
U_{(4)} &=& U_2(1.11)U_2(0.27)U_2(0.90)U_2(0.16)U_2(0.62),
\end{eqnarray}
whose optimal parameters are searched by sequential least squares programming.

\section{Deep Reinforcement Learning}

\label{DRL}

\begin{figure*}
\includegraphics[width=17.2cm]{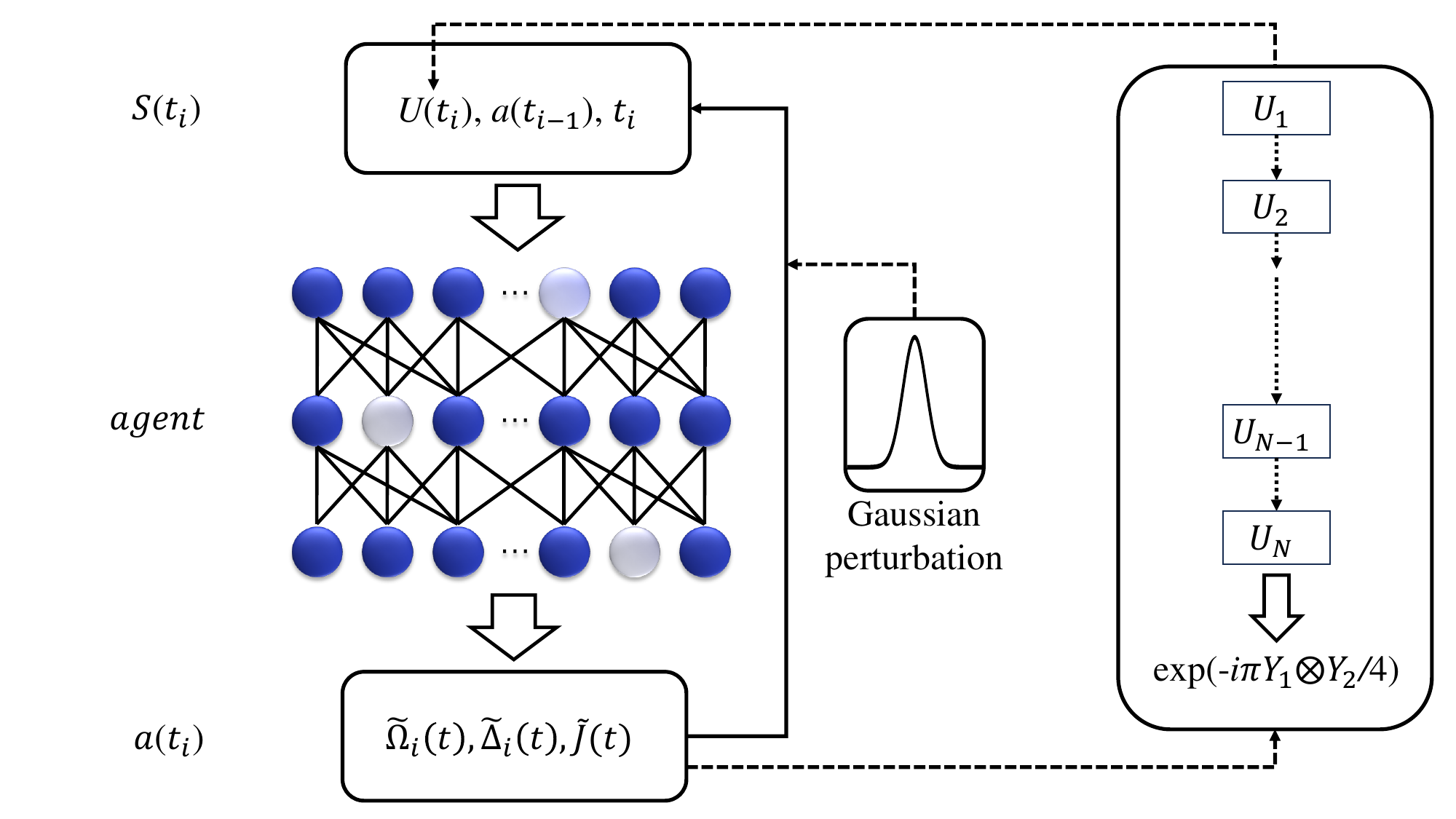}
\caption{\label{fig:workflow}Schematic diagram of deep reinforcement learning (DRL) approach to robust two-qubit gate. The state encodes the real part and imaginary part of elements in the propagator $U(t_i)$, the last action $a(t_{i-1})$, and the normalized systematic time $t_i=i/N_{\max}$. The state is sent to the input layer of the artificial neural network (ANN) as the DRL agent, outputting action to control the system for the next timestep. The agent is trained to accumulate maximum reward, targeting the two-qubit gate $R_{YY}(\pi/4)=\exp(i\pi Y_1\otimes Y_2/4)$ at the end of each episode. In our protocol, we use Gaussian perturbation on the action and dropout in the ANN to obtain robustness against systematic errors.}
\end{figure*}

After the analytical approach, we will move to deep reinforcement learning (DRL); as shown in its workflow (Fig.~\ref{fig:workflow}),  it is very similar to closed-loop quantum control. The insight is that a RL model can mimic the behavior of creatures that interact with the environment, being educated by reward to alter its decision based on its observation. In this way, an environment should be defined, specifying the task and its relevant state and action spaces, which can be either continuous or discrete. Next, an agent equipped with a deep neural network is trained to interact with the environment, making sequential decisions and receiving feedback in the form of rewards. Through an iterative process of exploration and exploitation, the agent refines its policy, using techniques like value-based methods or policy gradients, to maximize long-term cumulative rewards. After the training process, the agent can be deployed to perform the desired task autonomously, exhibiting learned behaviors that demonstrate the efficacy of DRL.

To be more specific, we aim to train a neural network as an agent for the robust design of the $R_{YY}(\pi/4)$ gate and benchmark the effects of physical priors or methods from the RL community that induce robustness. We unbound the constraints on $X_i$ and $Y_i$ in Eq.~\eqref{eqn:H}, resulting in the most general Hamiltonian for the agent to explore all possibilities
\begin{equation}
\label{eq:HDRL}
H_{DRL} = \frac{1}{2}\sum_{i=1}^2\left(\Omega_i(t)\cos\omega tX+\Omega_i(t)\sin\omega tY+\Delta_i(t)Z\right)+\frac{J(t)}{2}Z_1\otimes Z_2,
\end{equation}
where the tunable ranges of the parameters are $\Omega\in[0,\Omega_{\max}]$, $\Delta\in[-\Delta_{\max},\Delta_{\max}]$, and $J\in[0,J_{\max}]$. Accordingly, it is easy to calculate the renormalization of these control parameters to $[0,1]$ for ANN encoding in the action or state. Once we set the total operation time $T$, we obtain the length of each time step $\delta T=T/N_{\max}$ by bounding the maximum time steps per episode. At the $i$-th time step, the accumulated propagator reads
\begin{equation}
U(t_i,t_0)=\mathcal{T}\Pi_{j=0}^{i-1}U(t_{j+1},t_j)=\mathcal{T}\Pi_{j=0}^{i-1}\exp[-iH_{DRL}(j\delta T)\delta T],
\end{equation}
which should be close enough to $\exp(-i\pi Y_1\otimes Y_2)$ at the last time step after the training. The neural network is the general function approximator that denotes the policy $\pi(a|s)$, where the input layer observes the state $s(t_i)$, and the output layer provides the action $a(t_i)$ for evolving the state to the next time step. The state consists of the ordered elements of the propagator at the present, the last action, and the normalized systematic time as
\begin{equation}
s(t_i) = {\text{Re}U_{j,k}(t_i), \text{Im}U_{j,k}(t_i), a(t_{i-1}), \tilde{t}=i/N_{\max}},~j,k\in{1,2,3,4},
\end{equation}
where the action contains the renormalized control parameters in Eq.~\eqref{eq:HDRL} as
\begin{equation}
a(t_i)={\tilde{\Omega}_1(t_i), \tilde{\Delta}_1(t_i), \tilde{\Omega}_2(t_i), \tilde{\Delta}_2(t_i), \tilde{J}(t_i)}.
\end{equation}
The performance and training of RL are highly related to the reward function or pre-training. For a fair comparison among all models and to avoid cherry-picking, we do not perform any pre-training and define a very simple reward function by rewarding the agent with $r=-\log_{10}(1-F)$ in terms of logarithmic infidelity at the end of each episode, where the gate fidelity is defined as
\begin{equation}
F = \left|\frac{\text{Tr}[\exp(-i\pi Y_1\otimes Y_2/4)U(T,0)]}{\dim(U)}\right|^2.
\end{equation}
Using the baseline Proximal Policy Optimization (PPO) algorithm~\cite{schulman2017}, we train the agent to maximize the accumulated reward,  aiming  for leading to the quantum gates we seek.

\section{Numerical experiments}

\label{Numerical}

In this section, we present the specific settings for our numerical experiments and subsequently showcase our findings while analyzing the gate fidelities. Initially, we employ the most general Hamiltonian depicted in~\eqref{eq:HDRL} to evolve the quantum states, with the primary objective of assessing the capabilities of DRL in scenarios with limited prior physical knowledge.

\begin{figure*}
\includegraphics[width=17.2cm]{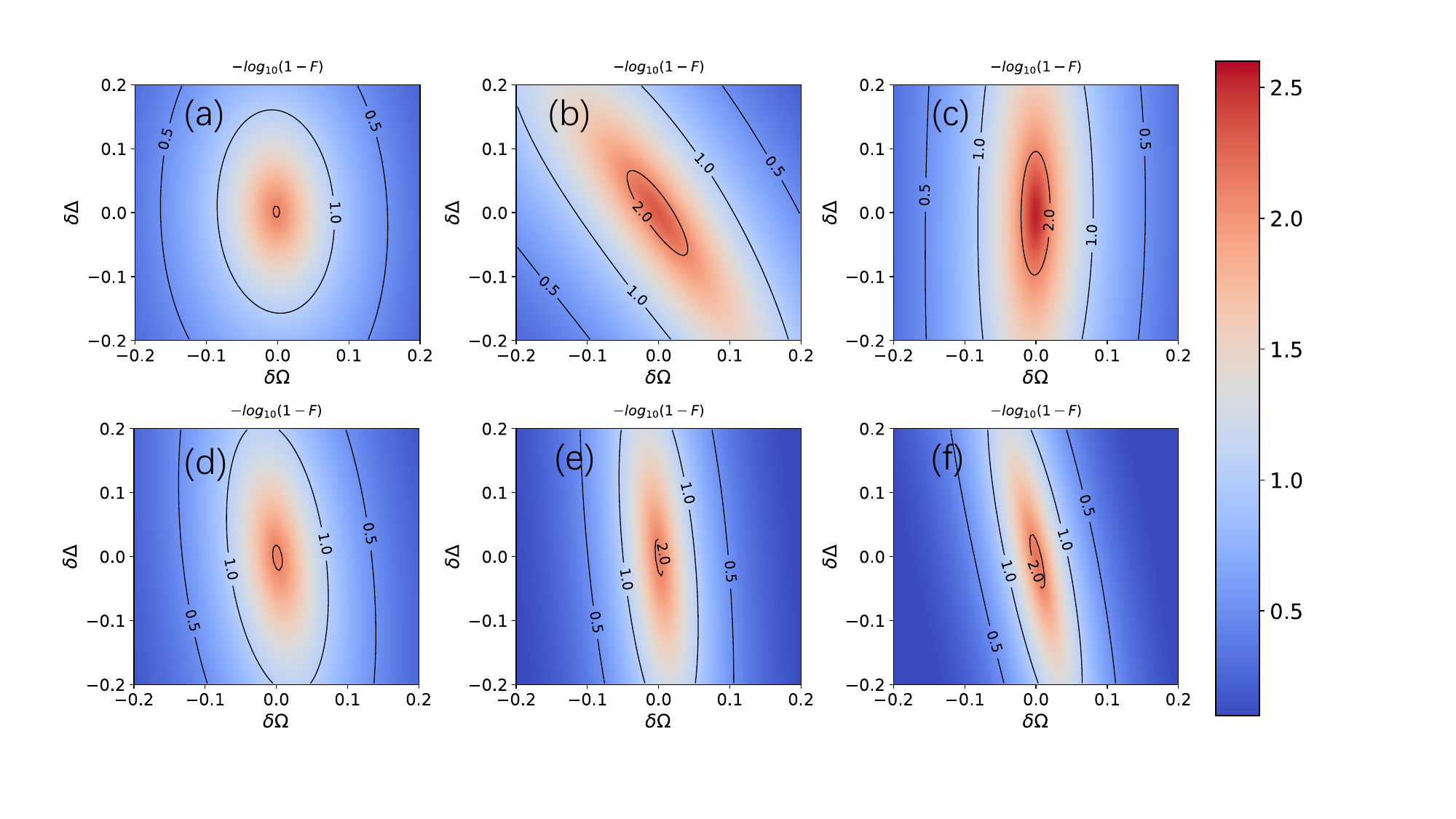}
\caption{\label{fig:heatmap} Robustness of the entangling gate $R_{YY}(\pi/4)$ against over-rotating errors ($\Omega_{1,2}\rightarrow\Omega_{1,2}+\Omega_{\max}\delta\Omega$) and off-resonance errors ($\Delta_{1,2}\rightarrow\Delta_{1,2}+\Delta_{\max}\delta\Delta$). The gate fidelity is defined as $F = \left|\frac{1}{4}\text{Tr}[\exp(-i\pi Y_1\otimes Y_2/4)U(T,0)]\right|^2$, where $U(T,0)$ represents the propagator under these errors. To emphasize the impact of systematic errors, we present a heatmap displaying the logarithmic infidelity. Models in (a-c) are trained in an environment featuring the general two-qubit Hamiltonian depicted in~\eqref{eq:HDRL}. These models are rewarded with logarithmic infidelity at the end of each episode, and they operate without prior human expert knowledge about robustness. We employ three different techniques from computer science: the standard batch method in (a), Gaussian perturbation on the nodes in the output layer in (b), and dropout in (c). Conversely, in (d), (e), and (f), we encode the Hamiltonian of the geometric gate shown in~\eqref{eqn:Hgeo}, which is based on explicit solutions, incorporating prior knowledge about the operation time, switching time, and Hamiltonian. Parameters: random Gaussian perturbations on nodes following $\tilde{\Omega},\tilde{\Delta}(t_i)\rightarrow \tilde{\Omega},\tilde{\Delta}[1+N(\mu=0,\sigma=0.1)]$, and a dropout rate of $\kappa=0.1$.}
\end{figure*}

The first setting is designed to efficiently obtain the gate operation. In this setting, we reward the agent with $r=-\log_{10}(1-F)$ after each time step in one episode, and conditionally end the episode to calculate the total reward if the maximum time step is met or $F>0.99$ is satisfied after any time step. Together with the discount rate $\gamma$ on the reward $\gamma^ir(t_i)$, this design encourages the agent to achieve the target more quickly by placing higher value on short-term rewards. Meanwhile, we clip the reward function by a value of $1$ if the gate fidelity falls within the range $[0.95,0.99)$, and provide an additional bonus of $10$ if it reaches the threshold of $F=0.99$. The clipped reward function can expedite training with the PPO algorithm without affecting our objectives. We halt the training and evaluate the model once it exceeds the fidelity threshold.

In Fig.~\ref{fig:heatmap}(a), we illustrate the gate's robustness against over-rotating errors ($\Omega_{1,2}\rightarrow\Omega_{1,2}+\Omega_{\max}\delta\Omega$) and off-resonance errors ($\Delta_{1,2}\rightarrow\Delta_{1,2}+\Delta_{\max}\delta\Delta$) under such a setting using a heatmap. As expected, the region enclosed by the contour $F=0.99$ is relatively small. However, this solution is not the time-optimal solution with minimal robustness due to the batch method employed in training the DRL agent. It can be easily verified that a time-optimal solution consists of single-qubit resonant pulses that exchange $Y$ and $Z$ and a two-qubit pulse of $ZZ$. In other words, DRL naturally introduces robustness into quantum control through the batch method,  which is a commonly used technique in ML, being applicable to gradient-based optimization as well~\cite{kang2021,heimann2023}. This result serves as the baseline for robustness to study the effects of other methods and settings.

\subsection{Gaussian perturbation on nodes}
To enhance robustness, we engineer the environment illustrated in Fig.~\ref{fig:heatmap}(a) with minimal human expert input. Instead of modifying the reward function by introducing Lagrangian multipliers for error sensitivity, we investigate whether perturbations applied to nodes of the ANN can achieve a similar effect. These perturbations are exclusively imposed on nodes within the output layer due to the inherent lack of interpretability of the network. The underlying idea is that if the DRL agent can be trained effectively in such an environment, it should exhibit greater robustness against over-rotating and off-resonance errors, as the environment simulates their effects by perturbing the agent's actions. In essence, we expect the agent to accumulate more rewards by successfully navigating this customized challenge.

We introduce perturbations to the action nodes using random Gaussian variables: $\tilde{\Omega},\tilde{\Delta}(t_i)\rightarrow \tilde{\Omega},\tilde{\Delta}[1+N(\mu=0,\sigma=0.1)]$. It is worthwhile to mention that we performed miscellaneous numerical experiments with different setting of standard deviation $\sigma=0.02, 0.05, 0.1, 0.2$, and observed the trade-off between the convergence of the model and robustness. The less the standard deviation is, the easier it is to obtain a converged model but the robustness is also reduced, and vice versa. Thus, we selected the standard deviation $\sigma=0.1$ as a trade-off. Notably, we observed that the DRL agent converged, as evidenced by the activation of the additional bonus once the gate fidelity surpasses the threshold under these perturbations. Consequently, we obtained control pulses by running the model in an error-free environment. Although the corresponding gate fidelity does not exceed $F=0.99$ due to the model converging to a solution with systematic errors induced by statistical fluctuations and the batch method, the characteristic of robustness remains intact.

In Fig.~\ref{fig:heatmap}(b), we rectify the deviated model by identifying the maximum fidelity point on the original heatmap and adjusting the control parameters to relocate the maximum fidelity to the center $(\delta\Omega,\delta\Delta)=(0,0)$ for illustrative purposes. This adjustment reveals a significantly larger area enclosed by the contour of $F=0.99$, demonstrating that Gaussian perturbations applied to nodes provide additional robustness in the design of quantum control using DRL. Moreover, the principal axis of the ellipse aligns with the diagonal direction because both types of systematic errors were included during the training process.

\subsection{Dropout}
Dropout is a common technique to mitigate overfitting during the training of ANNs~\cite{srivastava2014}. Larger weights in the network tend to overfit the training data more easily. The concept behind dropout is that probabilistically disconnecting nodes in the network serves as a simple regularization method to reduce weight magnitudes and perform model averaging. While this method was originally proposed for classical supervised learning, physicists have suggested its quantum analogue~\cite{wang2023} to reduce circuit complexity in quantum algorithms, aligning with the technique's underlying philosophy. This insight motivates us to investigate its performance in quantum control with DRL. The key idea is that we can emulate various systematic errors by randomly disconnecting nodes during training. Although we lack specific information about the correlation between a particular node and robustness against over-rotating or off-resonance errors, we believe that averaging the network across weights and biases should enhance gate robustness. The training process should converge when the dropout rate is appropriately set based on the network size and connectivity.

Instead of applying Gaussian perturbations to the output layer, we randomly disconnect each node with a dropout rate of $\kappa=0.1$. The setting of $\kappa$ is empirical and optimized, which has the similar effect on the model convergence and robustness as the standard deviation of Gaussian perturbation. While the ANN remains a black box, the lack of interpretability of specific nodes does not hinder our approach. Once the model converges, we disable dropout to obtain the control pulse, which is expected to be robust against all types of errors in the output layer, including variations in the length/magnitude of $ZZ$ interactions. After centralizing the heatmap using the same method, we achieve the best gate performance, as depicted in Fig.~\ref{fig:heatmap}(c). Based on the results of all our experiments, we conclude that DRL can explore solutions with robustness without the need for human expert intervention, relying instead on techniques from computer science. Particularly noteworthy is the fact that robustness can be introduced simply by implementing dropout, a regularization method, rather than modeling systematic errors in the environment or encoding criteria for robustness, such as error amplitudes or fidelity sensitivity, into the reward function.

\begin{figure*}
\includegraphics[width=17.2cm]{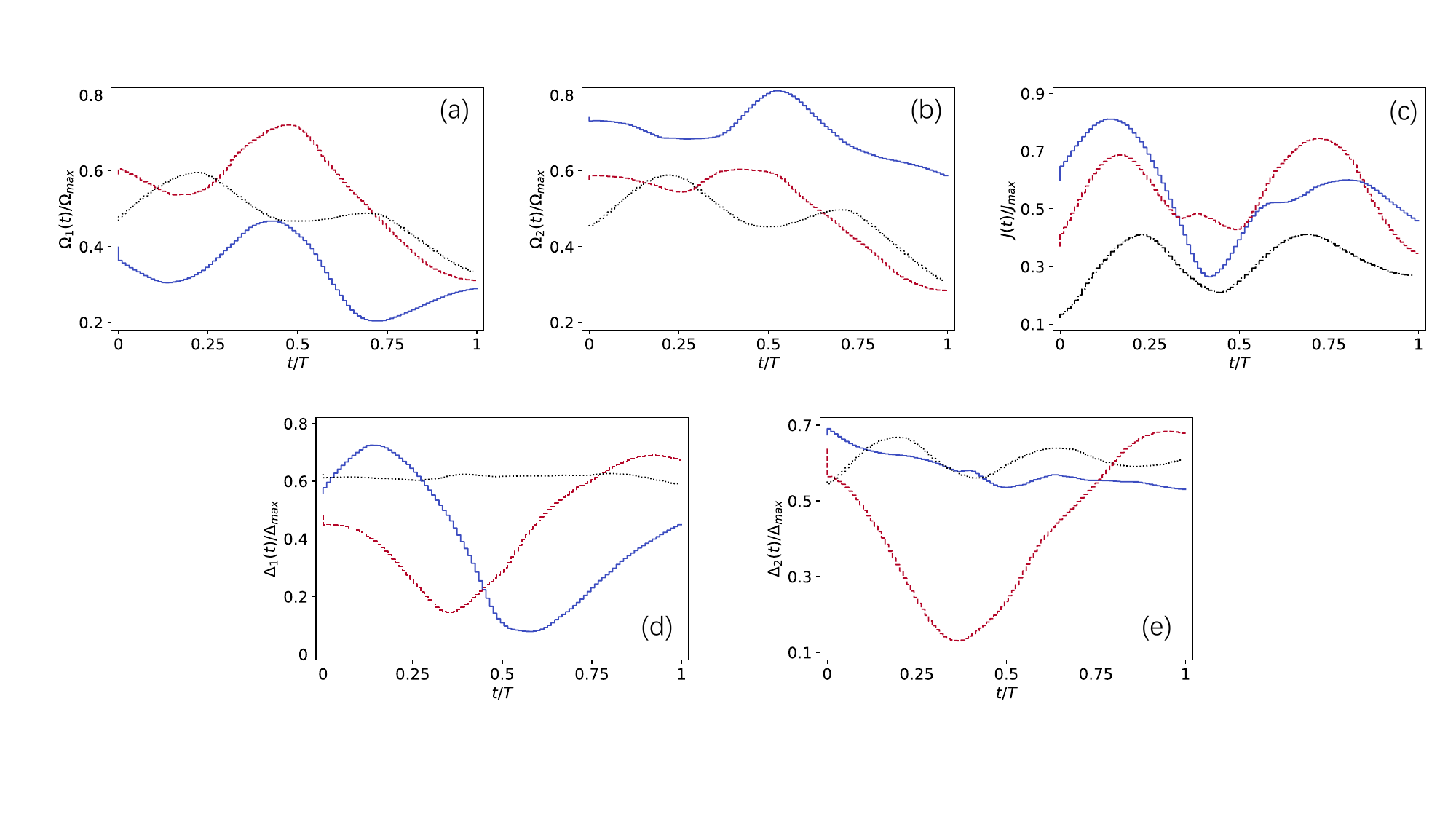}
\caption{\label{fig:pulse} Control pulses obtained from the models evaluated in Fig.~\ref{fig:heatmap}(a-c) without the guidance of human experts. (a) Rabi frequency on the first qubit. (b) Rabi frequency on the second qubit. (c) Magnitude of the $ZZ$ interaction between the two qubits. (d) Detuning on the first qubit. (e) Detuning on the second qubit. The solid blue lines represent pulses obtained from the standard batch method, the dashed red lines from Gaussian perturbation on nodes, and the dot-dashed black lines from dropout. Parameters: dimensionless tunable Rabi frequency $\Omega\in[0,\Omega_{\max}],~\Omega_{\max}=2\pi$, detuning $\Delta\in[-\Delta_{\max},\Delta_{\max}],~\Delta_{\max}=2\pi$, magnitude of $ZZ$ interaction  $J\in[0,J_{\max}],~J_{\max}=2\pi$, operation time $T=2$, maximal timestep $N_{\max}=100$.}
\end{figure*}

In Fig.~\ref{fig:pulse}, we present normalized control pulses for all tunable parameters explored using DRL. These pulses are obtained by running the trained DRL models in an error-free environment. It is remarkable that the pulses for single qubits are adjusted to centralize the maximum gate fidelity at $(\delta\Omega,\delta\Delta)=(0,0)$ within the corresponding parameter space. As outputs of the DRL, the pulse amplitudes are continuous in the decision layer but are discretized into stepwise functions, making them interpretable and suitable for smoothing if needed. We emphasize that these control pulses are applicable in most physical systems. Implementing the robust two-qubit gate involves calculating the operation time based on the tunable parameter ranges. In some specific cases, such as the weak coupling limit between qubits, one may need to retrain the model with modified tunable coupling ranges, but the methodology remains applicable in principle. Meanwhile, pulses should be smooth because actions taken in consecutive timesteps often exhibit notable similarities in PPO algorithm. This phenomenon underscores the temporal consistency of the policy, providing insights into the stability and predictability of the agent's decision-making process.

\subsection{Knowledge from geometric gate}
After verifying the capability of DRL in searching for robust gates without the guidance from human experts, we shift our focus to investigate if knowledge from geometric gates can inspire the agent to achieve better performance. Instead of the most general form of the Hamiltonian~\eqref{eq:HDRL}, we try to search the pulses characterized by the simplified two-qubit Hamiltonian~\eqref{eq:Hpm}, which resulted in geometric gates with explicit solutions [see Eq.~\eqref{eq:VU} and \eqref{eq:Ui}]. Based on the knowledge we gained from the analysis and optimization, we require the agent to run until it meets the maximal timestep in each episode. After bounding the total evolution time, we encode the gate time in the environment by evolving reduced single qubit Hamiltonian or two-qubit Hamiltonian~\eqref{eq:Hpm} as priori
\begin{widetext}
\begin{eqnarray}
\label{eqn:Hgeo}
H_{geo} &=& \sum_{i=1,2}\frac{1}{2}(\Omega_i(t)\cos\omega tX_i + \Omega_i(t)\sin\omega tY_i+\Delta(t) Z_i),\nonumber\\
&~&~~~~0<t<\max(T(U_{(1)}),T(U_{(2)}))\nonumber\\
&=& H^\pm(t) = \frac{1}{2}(\Omega(t)\cos\omega tG_x^\pm + \Omega(t)\sin\omega tG_y^\pm+\Delta^\pm(t) G_z^\pm),\nonumber\\
&~&~~~~\max(T(U_{(1)}),T(U_{(2)}))<t<\max(T(U_{(1)}),T(U_{(2)})+T(V_U)\nonumber\\
&=& \sum_{i=1,2}\frac{1}{2}(\Omega_i(t)\cos\omega tX_i + \Omega_i(t)\sin\omega tY_i+\Delta(t) Z_i),\\
&~&~~~~\max(T(U_{(1)}),T(U_{(2)})+T(V_U)<t<\max(T(U_{(1)}),T(U_{(2)})+T(V_U)+\max(T(U_{(3)}),T(U_{(4)})\nonumber.
\end{eqnarray}
\end{widetext}
Correspondingly, we modify our reward strategy, rewarding the agent with $r=-\log_{10}(1-F)$ only at the end of each episode, and we retain the clipping and extra bonus mechanisms to expedite training. Once we establish the tunable ranges $\Omega\in[0,\Omega_{\max}]$, $\Delta\in[-\Delta_{\max},\Delta_{\max}]$, and $J\in[0,J_{\max}]$, the Trotterized geometric $R_{YY}(\pi/4)$ gate becomes a part of the parameter space that can be explored. Our objective is to determine if the PPO algorithm, our revised reward functions, and techniques from computer science can lead to a solution when utilizing the standard batch method, Gaussian perturbations on the output layer, and dropout.

In Fig.~\ref{fig:heatmap}(d), (e), and (f), we observe that all these methods demonstrate a degree of robustness, but none outperforms the results obtained from the most general Hamiltonian combined with dropout. To delve deeper into the performance, we calculate the dynamical phases accumulated during the process, revealing that they are not canceled out. In other words, the mechanism responsible for robustness does not arise from the topological protection properties associated with geometric gates. Consequently, the model converges to a local maximum in terms of accumulated reward, resulting in pulses that exhibit lower robustness compared to the Trotterized geometric gate solution. Hence, our inclusion of the Hamiltonian and switching time as prior knowledge hinders the agent from exploring superior solutions when dealing with a more general form of Hamiltonian.

\section{Discussion}

\label{discussion}

This section delves into the more technical aspects after conducting a thorough analysis and evaluation of DRL's performance based on our previous numerical experiments. As one might discern, rewarding the agent with logarithmic infidelity at the end of each episode can inspire the agent to discover more efficient control pulses. In essence, it motivates the agent to explore an 'as-soon-as-possible' solution, aiming to achieve the gate fidelity threshold by the $N$-th timestep and conclude the episode before reaching the maximum timestep $N_{\max}$. We use "as-soon-as-possible" instead of "time-optimal" in the description, as an explicit objective function that encompasses terms related to robustness or energy cost is not provided in the problem statement. Notably, robustness in our approach is introduced through techniques from computer science, such as the batch method, perturbations on nodes, or dropout, rather than by engineering the reward function to incorporate error sensitivity.

Another critical consideration is that the DRL agent should not seek the global optimal solution that maximizes accumulated rewards in each episode.  For example,  in a scenario in which the agent can learn a gate operation with fidelity surpassing $0.99$ within a maximal timestep $N<N_{\max}$, it could exploit a loophole to gain extra rewards by deactivating all pulses once the clipped reward $r=1-\log_{10}(1-F)$ is triggered. The episode would continue since the gate fidelity does not meet the threshold. Finally, the agent could reactivate the pulses $a(t_N)$ at the last timestep of the episode to trigger the $r=10-\log_{10}(1-F)$ bonus. Through this method, the agent would earn an additional reward of
\begin{eqnarray}
\Delta r&=&\sum_{i=N-1}^{N_{max}-1}\gamma^{i}[1-\log_{10}(1-F_{N-1})]\nonumber\\
&~&+\gamma^{N_{max}}[10-\log_{10}(1-F_{N_{\max}})]\nonumber\\
&~&-\gamma^{N}[10-\log_{10}(1-F_{N_{\max}})],
\end{eqnarray}
which is positive when $N$ is sufficiently small and requires large tunable ranges of control parameters. This would result in a sudden drop in the total reward during the training process, even if the model appears to be converging. In our numerical experiments, neither of these phenomena occurred because we set the tunable ranges of control parameters within reasonable bounds, preventing the model from initially converging to an 'as-soon-as-possible' solution. However,  it should be noted that such solutions, as well as the cheating solution described, could potentially be discovered with certain hyperparameters after training for more episodes. In essence, what we performed in the numerical experiments, without guidance from human experts, corresponds to the early-stopping method,  a common technique to prevent overfitting in supervised learning.

The absence of geometric gates as robust solutions within specialized Hamiltonians in the environment is related to the initialization and exploration of the agent. None of the techniques from computer science led to geometric gates (at least not after training a moderate number of episodes) because the mechanism of robustness is not restricted to the property of topological protection. Our assumption is that DRL algorithms can discover geometric gates under settings with well-tuned hyperparameters and a more extensive number of episodes. Considering that the robustness performance shown in Fig.~\ref{fig:heatmap}(d), (e), and (f) is already quite promising compared to geometric gates, extensive effort would be required to uncover such solutions. However, we provide a trick to obtain geometric gates from the agent if one insists on pursuing this approach. One can evaluate the dynamical phases $\gamma_n^d$ at the end of each episode and design a reward function as follows: $r=-\log_{10}(1-F)-\lambda(\gamma_n^d)^2$, where $\lambda>0$ is a tunable coefficient for the Lagrangian multiplier. This additional term penalizes the agent by the square of dynamical phases if they are not canceled. Nevertheless, the requirement for this additional physical knowledge may discourage us from employing DRL for such a task, especially when the calculation of dynamical phases has already led to explicit solutions.

\section{Conclusion}

\label{conclusion}

We have investigated the capability of DRL in exploring robust two-qubit gates without the guidance of human experts. Through extensive numerical experiments, we have determined that DRL is capable of exploring robust quantum control using techniques from computer science. Specifically, dropout, which randomly disconnects nodes in the DRL agent during the training process, is found to provide satisfactory levels of robustness. It has become evident that there is no need to incorporate additional physical knowledge gained from analytical solutions, such as the operation time, switching time, or the form of the Hamiltonian. In fact, such additional information can hinder the agent from discovering the desired solutions. In essence, a general Hamiltonian with appropriately set tunable ranges and the incorporation of dropout is all that is required for the exploration of robust quantum control using DRL.

\begin{acknowledgements}
This work is supported by NSFC (12075145), STCSM (Grants No. 2019SHZDZX01-ZX04),  EU FET Open Grant EPIQUS (899368), HORIZON-CL4-2022-QUANTUM-01-SGA project 101113946 OpenSuperQPlus100 of the EU Flagship on Quantum Technologies, the Basque Government through Grant No. IT1470-22, the project grant PID2021-126273NB-I00 funded by MCIN/AEI/10.13039/501100011033, by ``ERDFA way of making Europe",  ``ERDF Invest in your Future", Nanoscale NMR and complex systems (PID2021-126694NB-C21),  the Valencian Government Grant with Reference Number CIAICO/2021/184, the Spanish Ministry of Economic Affairs and Digital Transformation through the QUANTUM ENIA project call -- Quantum Spain project, and the European Union through the Recovery, Transformation and Resilience Plan--NextGenerationEU within the framework of the Digital Spain 2026 Agenda.  X.C. acknowledges ``Ayudas para contratos Ram\'on y Cajal'' 2015-2020 (RYC-2017-22482).
\end{acknowledgements}

\clearpage
\begin{appendix}
\section{Hyperparamters}

\begin{table*}[!ht]
	\centering
	\caption{Hyperparameters of models in Fig.~\ref{fig:heatmap}}
	\begin{tabular}{|c|cccccc|}
		\hline
		~ & ~ & ~ & General Hamiltonian & ~ & ~  \\ \hline
		Subfigure & batch size & learning rate & episodes & operation time &maximal timestep & complementary  \\ \hline
		(a) & 64 & $10^{-4}$ & $10^5$ & 2 & 100 & default  \\ \hline
		(b) & 128 & $10^{-4}$ & $2\times10^5$ & 2 & 100 & $\tilde{\Omega},\tilde{\Delta}(t_i)\rightarrow \tilde{\Omega},\tilde{\Delta}[1+N(\mu=0,\sigma=0.1)]$  \\ \hline
		(c) & 128 & $10^{-4}$ & $2\times10^5$ & 2 & 100 & dropout rate $\kappa=0.1$  \\ \hline
		 ~ & ~ & ~ & Geometrical priori & ~ & ~ \\ \hline
		Subfigure & batch size& learning rate & episodes & operation time &maximal timestep & complementary \\ \hline
		(d) & 128 & $10^{-4}$ & $10^5$ & 17.05 & 100 & default \\ \hline
		(e) & 128 & $10^{-4}$ & $10^5$ & 17.05 & 100 & $\tilde{\Omega},\tilde{\Delta}(t_i)\rightarrow \tilde{\Omega},\tilde{\Delta}[1+N(\mu=0,\sigma=0.1)]$ \\ \hline
		(f) & 128 & $10^{-4}$ & $2\times10^5$ & 17.05 & 100 & dropout rate $\kappa=0.1$ \\ \hline
\end{tabular}
\label{datset}
\end{table*}

Besides these information for reproduction, we uploaded all codes, including environments, models, and scripts for evaluation to \href{https://github.com/wudiniuniu/DRL-for-two-qubit-gate}{GitHub}. Other hyperparameters of the PPO algorithm are set to be default if they are not mentioned. We used the open-source library TensorForce v.5.3.0 for the implementation.

\end{appendix}

\end{document}